\documentclass[review]{elsarticle}

\usepackage{lineno,hyperref}
\modulolinenumbers[5]

\journal{Journal of \LaTeX\ Templates}

\begin{document}

\begin{frontmatter}

\title{The ALICE Run 3 Online / Offline Processing}

\author{David Rohr for the ALICE Collaboration}
\address{CERN, 1 Esplanade des Particules, 1211 Geneva 23, Switzerland}
\ead{drohr@cern.ch}

\begin{abstract}
The ALICE experiment has undergone a major upgrade for LHC Run 3 and will collect data at an interaction rate 50 times larger than before.
The new computing scheme for Run 3 replaces the traditionally separate online and offline frameworks by a unified one, which is called O$^2$.
Processing will happen in two phases.
During data taking, a synchronous processing phase performs data compression, calibration, and quality control on the online computing farm.
The output is stored on an onsite disk buffer.
When there is no beam in the LHC, the same computing farm is used for the asynchronous reprocessing of the data which yields the final reconstruction output.
The O$^2$ project consists of three main parts:.
The Event Processing Nodes (EPN) equipped with GPUs deliver the bulk of the computing capacity and perform the majority of the reconstruction and the calibration.
The First Level Processors (FLP) receive the data via optical links from the detectors and perform local processing where it is needed, which can optionally happen in the user logic of the FPGA based readout card.
Between the FLP and the EPN farms the data is distributed in the network such that the EPNs receive complete collision data for the processing.
The Physics and Data Processing (PDP) group develops the software framework and the reconstruction and calibration algorithms.
The current O$^2$ setup is capable of handling in real time the peak data rate foreseen for data taking of Pb--Pb collisions at 50 kHz interaction rate..
\end{abstract}


\end{frontmatter}

ALICE (A Large Ion Collider Experiment) \cite{bib:alice} is the experiment at LHC dedicated to heavy ion collisions.
During the LHC long shutdown 2, ALICE underwent a major upgrade~\cite{bib:aliceupgrade} of several of its subdetectors and of the computing infrastructure.
The computing scheme splits the data processing into a synchronous stage during data taking performing data compression and calibration, and an asynchronous stage when there is no beam in the LHC.
The asynchronous processing runs partly on the online computing farm and partly on the computing GRID.

\section{Hardware and data flow}

The hardware consists of two computing farms with around 200 and 250 servers.
The FLP (First Level Processors) farm has FPGA based readout cards called CRUs (Common Readout Units), which receive the data via optical links from the detectors.
The FLP is the only place in the processing, where all data of an individual link is accessible.
Certain calibration operations, such as the integration of the digital currents of the Time Projection Chamber (TPC),  need access to all this data and must thus run on the FLPs.
Some detectors also run custom algorithms on the FPGA in the CRU user logic, e.\,g.~for Zero-Suppression of the raw data.

The data is transferred from the FLPs to the EPNs (Event Processing Nodes) via the InfiniBand network.
The transfer is arranged in such a way that an EPN server receives always the full data for individual collisions, while the different collisions are distributed over the EPNs.
The majority of the data processing is done on the EPNs.
Since TPC calibration and data compression rely on the information from TPC tracking, the EPNs perform full online track reconstruction of the TPC data.
This is the most computing-intense part of the synchronous processing.
Each EPN server is equipped with eight GPUs, which provide the bulk of the processing power.
The full TPC processing happens on the GPUs.

\section{Software}

The unified O$^2$ software framework is used for reconstruction and calibration both online and offline.
There are no separate algorithms, but the same framework and the same implementations are used for online and offline processing.
The online processing employs stricter cuts and skips some slow processing steps not needed for the compression or the calibration.
Since in Run 3 all data is processed online, not only collisions triggered by a hardware trigger, the computing requirements grow tremendously and ALICE and EPN use GPUs to speed up the processing~\cite{bib:gpu}.
The GPU and the CPU implementations share a common source code.
This significantly improves the maintainability, since only a single source code must be developed and maintained.

On top of processing, calibration, and data compression, the quality control (QC) performs a real time validation of the detector data on the EPNs, the FLPs, and dedicated QC servers.

The main purpose of the EPN farm is the synchronous processing of the peak data rate during the 50 kHz Pb--Pb data taking.
The TPC is the dominant contributor to this computing load, thus the EPNs are optimized for the fastest possible TPC processing on the GPUs.
The capability of the EPNs to handle the expected load was verified in a full-system test, in which simulated events were injected at the level of the data distribution entering the EPN, and the EPN was performing the full synchronous processing with its GPU and CPU parts.

The LHC collides heavy ions only during few weeks per year, thus the EPN farm will actually run asynchronous processing for most of the time.
It is thus desirable to use the computing capacity of the GPUs also in the asynchronous processing, wo avoid wasting the majority of the computing capacity.
While during the ongoing commissioning the most important task is still the readiness for the synchronous processing, there is an ongoing campaign to offload as much as possible parts of the asynchronous processing to the GPU.
A promising candidate is the full tracking chain in the central barrel region, including the Inner Tracking System (ITS), the Transition Radiation Detector (TRD), and the Time of Flight detector (TOF).
This could be further extended to the vertexing.
Not all of these components can fully run on the GPU yet, but already now the fraction that can be offloaded to the GPU is more than 80\% of the CPU computing load, measured in the case that all processing steps run on the CPU.
It is nevertheless important to offload also small but intermediate steps to avoid repeated data transfer forth and back.
The relative contribution of the GPUs to the total EPN computing capacity is around 90\%, and it is likely that eventually more than 90\% of the asynchronous computing load will be offloaded to the GPU, such that the GPUs will be fully loaded in both computing phases.

The ALICE online and offline systems have proven their readiness for data taking during the LHC pilot beam test at the end of 2021 running full online processing with GPUs.
The next steps are the pp data taking at nominal interaction rate, the Pb--Pb run at the peak design interaction rate of 50\,kHz, and the asynchronous processing of the data.

\section*{Acknowledgements}

We thanks German BMBF and Greek ESPA 2014-2020 National Fund for Research Infrastructures DeTAnet for support of the Alice Offline/Online project.

\end{document}